\documentclass[twocolumn,showpacs,aps,prl,superscriptaddress]{revtex4}

\usepackage{graphicx}
\usepackage{dcolumn}
\usepackage{amsmath}

\input pubboard/babarsym

\def\psitwos {\ensuremath{{\psi(2S)}}}

\def\jpsiee {\ensuremath{{\jpsi\to\epem}}}
\def\jpsimm {\ensuremath{{\jpsi\to\mumu}}}
\def\psitwosee {\ensuremath{{\psitwos\to\epem}}}
\def\psitwosmm {\ensuremath{{\psitwos\to\mumu}}}
\def\psitwospp {\ensuremath{{\psitwos\to\jpsi\pipi}}}

\def\eeresult {\ensuremath{{0.0078 \pm 0.0009 \pm 0.0008}}}
\def\mmresult {\ensuremath{{0.0067 \pm 0.0008 \pm 0.0007}}}
\def\eebr {\ensuremath{{0.0252 \pm 0.0028 \pm 0.0011}}}
\def\mmbr {\ensuremath{{0.0216 \pm 0.0026 \pm 0.0014}}}
\def\mmeeratio {\ensuremath{{0.86 \pm 0.12 \pm 0.05}}}

\newcommand{\BABARPubYear}    {01}
\newcommand{\BABARPubNumber}  {13}

\newcommand{\SLACPubNumber} {8953}
\newcommand{\LANLNumber} {0109004}

\def\figurebox#1#2#3{%
    \def\arg{#3}%
    \ifx\arg\empty
    {\hfill\vbox{\hsize#2\hrule\hbox to #2{\vrule\hfill\vbox to #1{\hsize#2\vfill}\vrule}\hrule}\hfill}%
    \else
    {\hfill\epsfbox{#3}\hfill}%
    \fi}

\long\def\inst#1{\par\nobreak\kern 4pt\nobreak
    {\it #1}\par\vskip 10pt plus 3pt minus 3pt}

\begin{document}

\begin{flushleft}
%\babar\ Analysis Document \# 192, Version 06\\
\babar-PUB-\BABARPubYear/\BABARPubNumber\\
SLAC-PUB-\SLACPubNumber\\
hep-ex/\LANLNumber\\[10mm]
\end{flushleft}

\title{
 Measurement of the branching fractions for \psitwosee\ and \psitwosmm 
\begin{center}
\vskip 10mm
The \babar\ Collaboration
\end{center}
}

%% author list as of 01-Aug-2001 (588 authors)
%
\author{B.~Aubert}
\author{D.~Boutigny}
\author{J.-M.~Gaillard}
\author{A.~Hicheur}
\author{Y.~Karyotakis}
\author{J.~P.~Lees}
\author{P.~Robbe}
\author{V.~Tisserand}
\affiliation{Laboratoire de Physique des Particules, F-74941 Annecy-le-Vieux, France }
\author{A.~Palano}
\author{A.~Pompili}
\affiliation{Universit\`a di Bari, Dipartimento di Fisica and INFN, I-70126 Bari, Italy }
% ...
%\collaboration{The \babar\ Collaboration}
%\noaffiliation

\date{September 9, 2001}

\begin{abstract}
We measure the branching fractions of the \psitwos\ meson to the
leptonic final states \epem\ and \mumu\ relative to that for 
\psitwospp.
The method uses \psitwos\ mesons produced in the decay of $B$ mesons
at the \FourS\ resonance in a data sample collected with the \babar\
detector at the Stanford Linear Accelerator Center.  Using previous
measurements for the \psitwospp\ branching fraction, we determine the
\epem\ and \mumu\ branching fractions to be \eeresult\ and \mmresult\
respectively. 
\end{abstract}

% PACS, the Physics and Astronomy Classification Scheme.

\maketitle

The branching fraction of the \psitwos\ to \epem\ has previously been
measured in \epem\ collider experiments operating at the mass of the
\psitwos\ resonance \cite{ref:pdgee}
and in $p\bar p$ experiments \cite{ref:e760, ref:e835}.
The
$\psitwos\to\mumu$ branching fraction has been measured 
with substantially larger uncertainty in \epem\
experiments \cite{ref:pdgmm}
and in $\pi^- Be$ collisions \cite{ref:e672}.  
This paper reports
new measurements of these quantities by the \babar\ experiment, operating 
at the PEP-II \epem\ collider at the Stanford Linear Accelerator
Center.  

%These final states are of particular interest
%because of their use in reconstructing $B$ mesons for 
%\CP\ violation and branching fraction measurements
%\cite{ref:excl}. 

PEP-II collides $9$~GeV electrons on $3.1$~GeV positrons to create a
center-of-mass system with energy 10.58\gev\ moving along the $z$
axis with a Lorentz boost of $\beta\gamma = 0.56$.  At this energy,
\FourS\ resonance production makes up 23\% of the total hadronic cross
section.  The
\FourS\ is assumed to decay 100\% to a pair of $B$ mesons.
A large, clean sample of \psitwos\ mesons is produced in the $B$
decays.  The \epem\ and \mumu\ branching fractions are obtained
through their ratio to $\jpsi\pipi$, which is
known with much better precision.   This technique provides a 
significantly lower uncertainty on the \mumu\ branching fraction than
the current world average.

The data set used for this analysis corresponds to 
an integrated luminosity of $20.33\pm
0.30$\invfb\ recorded at 10.58\gev, and contains 
$\left( 22.3 \pm 0.4
\right )\times 10^6$ \FourS\ mesons.  An additional 2.6\invfb\ has
been recorded at an
energy 40\mev\ below the \FourS\ resonance.

The \babar\ detector is described in detail in
reference~\cite{ref:babarnim}.  
The momenta of charged particles are measured and their trajectories
reconstructed with two detector systems located in a
1.5-T solenoidal magnetic field: a five-layer, double-sided silicon
vertex tracker (SVT) and a 40 layer drift chamber (DCH).  The fiducial
volume covers the polar angular region $0.41 < \theta < 2.54$\,rad, 
which is 86\% of the
solid angle in the center of mass.  The transverse momentum resolution
is 0.47\% at 1\gevc.

The energies of electrons and photons are accurately measured by a
CsI(Tl) calorimeter (EMC) in the fiducial volume
$0.41 < \theta < 2.41$\,rad (84\% of the center-of-mass solid angle) 
with energy resolution at 1\gev\ of 3.0\%.
Muons are detected in
the IFR---the flux return of the solenoid, which is instrumented with
resistive 
plate chambers.   The DIRC, 
a unique Cherenkov radiation detection device, identifies
charged particles.  

The branching fractions of interest are obtained by comparison to that 
of \psitwospp.  
The number of \psitwos\ mesons reconstructed in the final
states \epem\ ($N_{ee}$), \mumu\ ($N_{\mu\mu}$) and 
$\jpsi\pipi$, with \jpsiee\ ($N_{ee\pi\pi}$) or \jpsimm
($N_{\mu\mu\pi\pi}$), is related to 
the total number of \psitwos\ mesons produced in our
data set $N_{\psitwos}$ by:
\begin{eqnarray}
N_{ee} & = & N_{\psitwos} \cdot \BR_{ee} 
  \cdot \epsilon_{ee}, \label{eqn:nee} \\
N_{\mu\mu} & = & N_{\psitwos} \cdot \BR_{\mu\mu} 
  \cdot \epsilon_{\mu\mu}, \label{eqn:nmm} \\
N_{ee\pi\pi} & = & N_{\psitwos} \cdot \BR_{\jpsi\pipi} \cdot 
  \BR_{\jpsi\to ee} \cdot \epsilon_{ee\pi\pi},
 \label{eqn:neepp}\\
%\text{and \hfill \ } & &  \nonumber \\ 
N_{\mu\mu\pi\pi} & = & N_{\psitwos} \cdot \BR_{\jpsi\pipi} \cdot 
  \BR_{\jpsi\to \mu\mu} \cdot \epsilon_{\mu\mu\pi\pi}.
 \label{eqn:nmmpp}
\end{eqnarray}
$\BR_{ee}$, $\BR_{\mu\mu}$ 
and $\BR_{\jpsi\pipi}$ are the branching fractions of the
\psitwos\ to \epem, \mumu, and $\jpsi\pipi$ respectively. 
We use world averages for $\BR_{\jpsi\to ee}$,
the \jpsi\ branching fraction to \epem, and for $\BR_{\jpsi\to
\mu\mu}$, the branching fraction to \mumu \cite{ref:pdg2000}.   
$\epsilon_{ee}$ and $\epsilon_{\mu\mu}$
are the 
efficiencies for events containing \psitwos\ mesons decaying to
\epem\ and \mumu\ respectively to satisfy the
event selection and meson reconstruction requirements; 
$\epsilon_{ee\pi\pi}$ and $\epsilon_{\mu\mu\pi\pi}$ are the
efficiencies for \psitwospp\ decays with \jpsiee\ and \jpsimm\
respectively. 

Equations \ref{eqn:nee}, \ref{eqn:neepp}, \ref{eqn:nmmpp}
can be combined to give two expressions for the
\epem\ to $\jpsi\pipi$ branching ratio :
\begin{eqnarray}
\frac{\BR_{ee}}{\BR_{\jpsi\pipi}} & = & \BR_{\jpsi\to ee} \cdot
\frac{N_{ee}}{N_{ee\pi\pi}} \cdot 
\frac{\epsilon_{ee\pi\pi}}{\epsilon_{ee}},
\label{eqn:breeee} \\
\frac{\BR_{ee}}{\BR_{\jpsi\pipi}} & = & \BR_{\jpsi\to\mu\mu} \cdot
\frac{N_{ee}}{N_{\mu\mu\pi\pi}} \cdot 
\frac{\epsilon_{\mu\mu\pi\pi}}{\epsilon_{ee}}.
\label{eqn:breemm}
\end{eqnarray}
Similarly, 
\begin{eqnarray}
\frac{\BR_{\mu\mu}}{\BR_{\jpsi\pipi}} & = & \BR_{\jpsi\to ee} \cdot
\frac{N_{\mu\mu}}{N_{ee\pi\pi}} \cdot 
\frac{\epsilon_{ee\pi\pi}}{\epsilon_{\mu\mu}},
\label{eqn:brmmee} \\
\frac{\BR_{\mu\mu}}{\BR_{\jpsi\pipi}} & = & \BR_{\jpsi\to\mu\mu} \cdot
\frac{N_{\mu\mu}}{N_{\mu\mu\pi\pi}} \cdot 
\frac{\epsilon_{\mu\mu\pi\pi}}{\epsilon_{\mu\mu}}.
\label{eqn:brmmmm}
\end{eqnarray}
A number of systematic errors due to uncertainties in efficiency
cancel in these expressions. 

We obtain a \BB\ enriched sample by requiring 
events to have visible energy $E$ greater than
4.5\gev\ and a ratio of the second to the zeroth Fox-Wolfram moment,
$R_2$
\cite{ref:fox}, less than 0.5.  Both $E$ and $R_2$ are calculated from 
tracks and neutral clusters in the respective fiducial volumes noted
above.  The same tracks are used to construct a primary event vertex,
which is required to be located within 6\cm\ of the beam spot in $z$
and within 0.5\cm\ of the beam line. The beam spot rms size is
approximately 0.9\cm\ in $z$, 120\mum\ horizontally, and 5.6\mum\
vertically.

There must be at least three tracks in the fiducial volume satisfying
the following quality criteria: they must have transverse momentum
greater than 0.1\gevc, momentum less than 10\gevc, at least 12 hits in
the DCH, and approach within 10\cm\ of the beam spot in $z$ and within
1.5\cm\ of the beam line.

Finally, to suppress a substantial background from radiative Bhabha
($\epem\gamma$) events in which the photon converts to an \epem\ pair,
five or more tracks are required in events containing \psitwosee\ or
\jpsiee\ candidates.

The efficiency of the event selection---and the meson reconstruction
efficiency described below---is calculated with a complete detector
simulation of $B\to\psitwos X$ events \cite{ref:geant}.  
The simulation of \psitwos\ and \jpsi\ decays to lepton pairs includes 
final state radiation \cite{ref:photos}.
The event
selection efficiencies are $0.912\pm 0.002$ for \psitwosee, $0.945 \pm
0.002$ for \psitwosmm, $0.967 \pm 0.001$ for $\epem\pipi$, and $0.972
\pm 0.001$ for $\mumu\pipi$.  The difference in the \epem\ and \mumu\
efficiencies is due largely to the requirement of five tracks.  The
quoted uncertainties are those due to simulation statistics only.  The
event efficiencies appear as ratios in 
equations~\ref{eqn:breeee}--\ref{eqn:brmmmm}; 
the systematic errors on the ratios are small
compared to the other uncertainties and systematic errors discussed
below.

The lepton candidates used to construct \jpsi\ or \psitwos\ mesons via
\epem\ or \mumu\ decays must be in the restricted angular region
$0.41 < \theta < 2.41$\,rad and satisfy 
the track quality criteria listed above.

Electron candidates must include an energy deposition in the EMC of at
least three crystals, with shape consistent with an electromagnetic
shower and magnitude at least 75\% of the track momentum.  At least
one candidate must have energy between 89\% and 120\% of the track
momentum and a Cherenkov signal in the DIRC consistent with the
expectation for an electron.
If possible,
photons radiated by electrons traversing material prior to the DCH are
recombined with the track.  Such photons must have EMC energy greater
than 30\mev, a polar angle $\theta$ within 35\mrad of the electron
direction and an azimuth that is either within 50\mrad of the electron 
direction or between the electron direction and the location of the 
electron shower in the EMC.

Muon candidates must deposit less than 0.5\gev\ in the EMC
(2.3 times the minimum-ionizing peak), 
penetrate at least
two interaction lengths $\lambda$ of material, and have a pattern of
hits consistent with the trajectory of a muon.
We require the material traversed by one candidate be within 
1\,$\lambda$ of that expected for a muon; for the other candidate, this is 
relaxed to 2\,$\lambda$.

The \jpsi\ or \psitwos\ meson mass is obtained in an \ellell\ final
state after constraining the two tracks to a common origin.

The reconstruction of $\psitwospp$ uses a
\jpsiee\ candidate with mass between 3.05 and 3.12\gevcc\ or a
\jpsimm\ candidate with $3.07 < m < 3.12$\gevcc.  74\% of \jpsiee\
decays and 91\% of \jpsimm\ fall within these ranges.  All tracks in
the fiducial volume not used in the \jpsi\ reconstruction are used as
pion candidates.  To avoid systematic errors and retain
high efficiency, 
the tracks are not required to satisfy any specific
quality requirements.
A pair of oppositely-charged pions is required to
have mass $m_{\pi\pi}$ in the region $0.45 < m_{\pi\pi} < 0.60$\gevcc.
The \psitwos\ mass is obtained after constraining the four tracks in
the final state to a common origin.

\psitwos\ candidates in all final states are required to have momentum 
less than 1.6\gevc\ as measured in the \FourS\ rest frame.  This
requirement is fully efficient for \psitwos\ mesons produced in $B$
decays.

%The average number of candidates in events with at least one 
%is 1.001 for $\psitwos\to\ellell$ and 1.8 for $\jpsi\pipi$.  All
%candidates are used in the analysis.

The \jpsi\ and \psitwos\ reconstruction efficiencies are determined by
simulation and include 
contributions from acceptance, track quality, particle identification
and, for \psitwospp, the \jpsi\ and \pipi\ mass windows.
The efficiency and systematic error on 
lepton identification have been obtained from data by
comparing the ratio of \jpsi\ mesons in $B$ decays in which one or
both leptons satisfy the requirements.  The efficiency and systematic
error of the
track-quality selection have been studied by comparing the independent
SVT and DCH tracking efficiencies in hadronic events.  The meson
reconstruction efficiency is $0.602\pm 0.004$ for the \epem\ case, 
$0.535\pm 0.004$ for
\mumu, $0.207 \pm 0.002$ for $\epem\pipi$, and $0.211 \pm 0.002$ 
for $\mumu\pipi$, where the uncertainties are simulation statistics
only.  

The \epem\ efficiency is higher than \mumu\ in $\psitwos\to\ellell$ or
$\jpsi\to\ellell$ reconstruction because electron identification is
more efficient than muon identification.  Conversely, a \jpsi\
decaying to \epem\ is less likely to be reconstructed in the specified
mass window than one decaying to \mumu.  Together, these two effects
result in little difference between the $\epem\pipi$ and $\mumu\pipi$
efficiencies.  Overall, the $\jpsi\pipi$ efficiencies are lower than
\ellell\ due to the reconstruction of the pion pair.  The efficiencies
appearing in equations \ref{eqn:nee}--\ref{eqn:nmmpp} are the product
of these meson reconstruction efficiencies and the event selection
values given earlier.

Lepton
identification uncertainty is 1.8\% for \epem\ and 1.4\% for \mumu,
and cancels in branching ratios where the \psitwos\ and \jpsi\ decay
to the same final state, equations~\ref{eqn:breeee} and
\ref{eqn:brmmmm}.   
A 2.4\% systematic error on the efficiency
of the track quality requirements applied to the \jpsi\ and \psitwos\
in the \ellell\ final state
cancels in all four ratios.

\begin{figure}
\includegraphics[width=\linewidth]{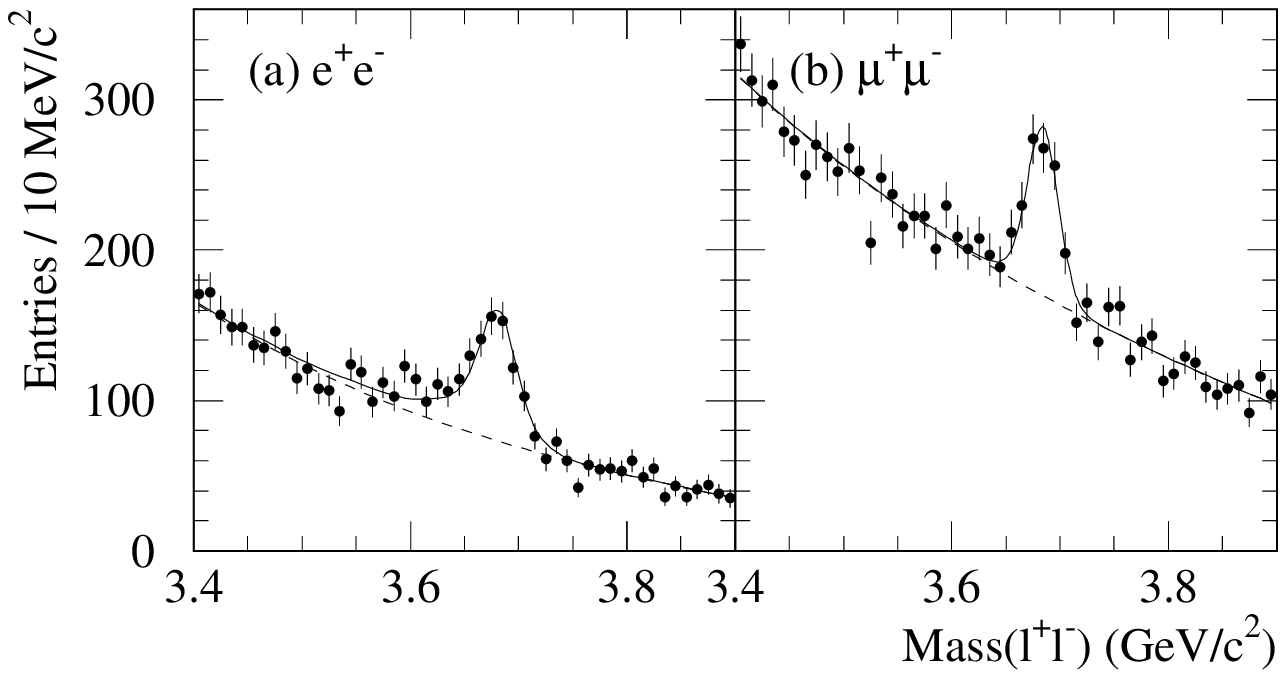}
\caption{Mass distribution of (a) \psitwosee\ and (b) \psitwosmm\
candidates.} 
\label{fig:ll}
\end{figure}

The number of mesons in the \epem\ and \mumu\ final states is
extracted by a fit to the mass distribution of candidates
(Fig.~\ref{fig:ll}).  A third-order Chebychev polynomial is used for
backgrounds.  The signals are fit by probability distribution
functions (pdfs) obtained from a complete simulation of $B\to \psitwos
X$ events, with \psitwosee\ or \psitwosmm.  Only candidates
constructed from the correct combination of particles are used in the
pdf.  The signal pdfs are convoluted with a Gaussian distribution to
match the mass resolution of 12\mevcc\ observed in a data sample of
14,000 \jpsimm\ decays.  

Despite the algorithm to recover radiated photons, the pdf for the
\epem\ final state is sensitive to the fraction of events in which one
or both electrons undergo bremsstrahlung.  The pdf is adjusted to
reflect the fraction obtained in a study of the mass distribution of
15,000 \jpsiee\ decays in data.  To enhance the sensitivity of the
study, the algorithm to recover radiated photons is not used in the
reconstruction of the \jpsi.

\begin{figure}
\includegraphics[width=\linewidth]{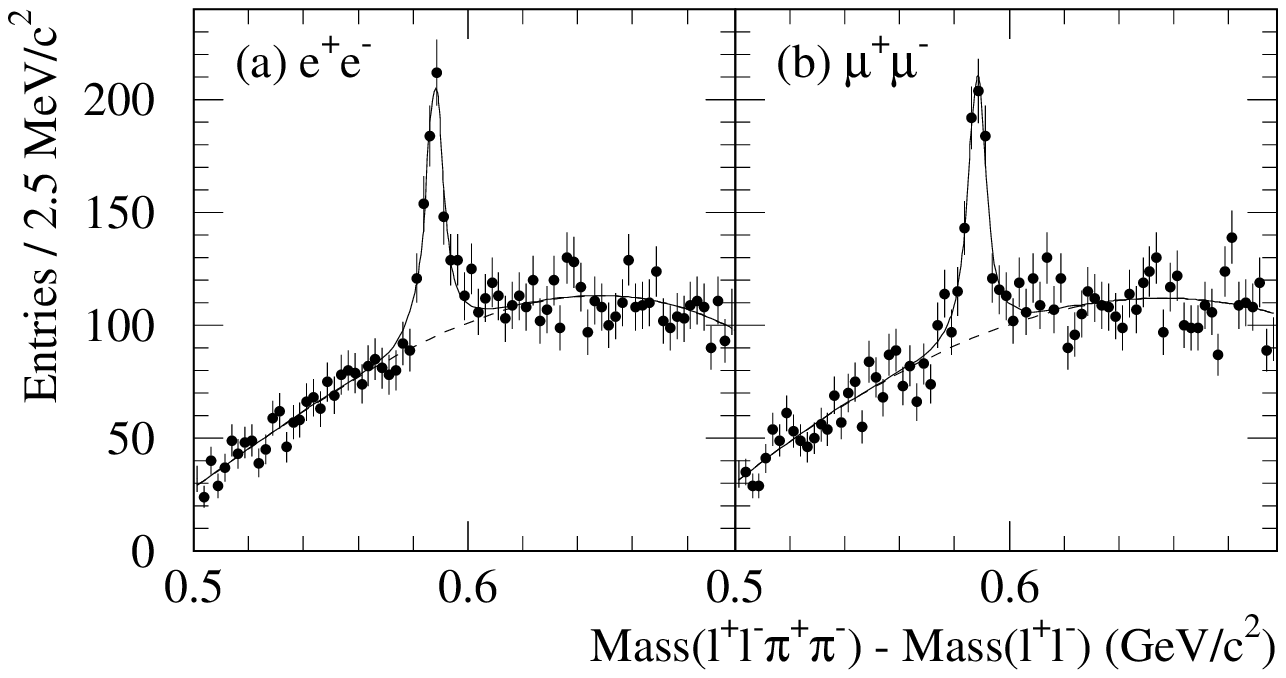}
\caption{Mass difference between the \psitwos\ and \jpsi\ candidates
in the decay \psitwospp\ with the \jpsi\ reconstructed in the (a)
\epem\ and (b) \mumu\ final states.}
\label{fig:llpp}
\end{figure}

For \psitwospp, an analogous fit procedure is performed to the
distribution of the mass difference between the \psitwos\ and the
\jpsi\ candidates (Fig.~\ref{fig:llpp}).
This quantity reduces the impact of \jpsi\ mass resolution, including
final state 
radiation and bremsstrahlung.  The distribution predicted by the
simulation is convoluted with a Gaussian distribution whose standard
deviation is left as a free parameter in the fit.
The mass difference resolution is 3.2\mevcc.  

The signal yields returned by the fits are $552 \pm 50$ for \epem,
$437 \pm 44$ for \mumu, $474 \pm 44$ for $\epem\pipi$, and $498\pm 42$
for $\mumu\pipi$, where errors are statistical only.  Systematic
errors on the fitting technique are obtained by performing the fits on
multiple simulated data sets containing both signal and background
events. 
Additional contributions come from varying the mass regions included
in the fit and increasing or decreasing the power of the background
polynomial.  Fitting systematics are 2.3\% for \epem, 5.3\% for \mumu,
5.4\% for $\epem\pipi$, and 2.1\% for $\mumu\pipi$.  These systematic
errors are conservative in the sense that the procedure to derive them
incorporates a component of the statistical error, which would be
reduced with additional data.

We repeat the analysis with the data recorded below the \FourS\
resonance.  The total \psitwos\ yield, summed over the four modes, is
$5\pm 12$ events, indicating that the contribution of
continuum-produced \psitwos\ mesons is negligible in the on-resonance
sample. 

The two values for the \epem\ to $\jpsi\pipi$ branching ratio obtained
with equations
\ref{eqn:breeee} and \ref{eqn:breemm} are in good agreement: the
result found with $\mumu\pipi$ is $0.97 \pm 0.14$ times that with
$\epem\pipi$.  By construction, this ratio is identical for the \mumu\
final state.  The results from equations \ref{eqn:breeee} and
\ref{eqn:breemm} are combined, 
distinguishing correlated and uncorrelated statistical and systematic
errors, to give:
\begin{equation}
\BR_{ee}/\BR_{\jpsi\pipi} =  \eebr,
\label{eqn:eebr}
\end{equation}
where the first error is statistical and the second systematic. 
Similarly, equations \ref{eqn:brmmee} and \ref{eqn:brmmmm} are
combined to obtain 
\begin{equation}
\BR_{\mu\mu}/\BR_{\jpsi\pipi}  =  \mmbr. \label{eqn:mmbr}
\end{equation}
The systematic errors are dominated by the
fitting technique.  Other contributions, which are the same for both
results, include 1.6\% for particle identification, 1.2\% for the
uncertainty in \jpsi\ branching fractions, and 0.9\% for differences
between the simulated and measured \cite{ref:bes}
\pipi\ mass and angular distributions in the $\jpsi\pipi$ final states.

We use the current world average value of $0.310 \pm 0.028$ for
the \psitwospp\ branching fraction \cite{ref:pdg2000} 
to extract results for the \psitwos\ 
leptonic branching fractions:
\begin{eqnarray}
\BR_{ee} & = & \eeresult, \label{eqn:eeresult} \\
\BR_{\mu\mu} & = & \mmresult. \label{eqn:mmresult}
\end{eqnarray}

The ratio of the leptonic branching fractions can be derived without
the use of the \psitwospp\ sample:
\begin{equation}
\frac{\BR_{\mu\mu}}{\BR_{ee}} = 
 \frac{N_{\mu\mu}}{N_{ee}} \cdot
\frac{\epsilon_{ee}}{\epsilon_{\mu\mu}} = \mmeeratio.
\end{equation}
The systematic error is dominated by the uncertainty in the fitting
technique.

In summary, we have measured the branching ratios
$\BR_{ee}/\BR_{\jpsi\pipi}$ and $\BR_{\mu\mu}/\BR_{\jpsi\pipi}$.  We
multiply these by the world average for the $\jpsi\pipi$ branching
fraction to obtain the branching fraction of the \psitwos\ to \epem\
and to \mumu.  These results are consistent with earlier measurements,
but have, in the case of \mumu, a substantially smaller uncertainty.

We are grateful for the excellent luminosity and machine conditions
provided by our \pep2\ colleagues, 
and for the substantial dedicated effort from
the computing organizations that support \babar.
The collaborating institutions wish to thank 
SLAC for its support and kind hospitality. 
This work is supported by
DOE
and NSF (USA),
NSERC (Canada),
IHEP (China),
CEA and
CNRS-IN2P3
(France),
BMBF and DFG
(Germany),
INFN (Italy),
NFR (Norway),
MIST (Russia), and
PPARC (United Kingdom). 
Individuals have received support from the 
A.~P.~Sloan Foundation, 
Research Corporation,
and Alexander von Humboldt Foundation.

\end{document}